\documentclass[9pt, a4paper, conference, compsocconf]{IEEEtran}

\usepackage{graphicx}
\usepackage{bmpsize}
\usepackage{epstopdf}
\usepackage{amsmath,amssymb,booktabs,tabularx}
\usepackage[hyphens]{url}
\usepackage{hyperref}
\def\UrlBreaks{\do\/\do-}
\usepackage{mathrsfs}
\usepackage{float}
\usepackage{multirow}
\usepackage{soul}
\usepackage[table]{xcolor}
\usepackage{comment}
\usepackage{subcaption}
\usepackage{mwe}
\usepackage{slashbox}
\usepackage{multicol}
\usepackage{import}
\usepackage{siunitx}
\usepackage{mathtools,url}
\usepackage{amssymb}
\usepackage{enumerate}
\usepackage{breakcites}
\PassOptionsToPackage{hyphens}{url}\usepackage{hyperref}
\makeatletter
\g@addto@macro{\UrlBreaks}{\UrlOrds}
\makeatother

\usepackage{makecell} 

\makeatletter
\let\old@ps@headings\ps@headings
\let\old@ps@IEEEtitlepagestyle\ps@IEEEtitlepagestyle
\def\confheader#1{%
	\def\ps@IEEEtitlepagestyle{%
		\old@ps@IEEEtitlepagestyle%
		\def\@oddhead{\strut\hfill#1\hfill\strut}%
		\def\@evenhead{\strut\hfill#1\hfill\strut}%
	}%
	\ps@headings%
}
\makeatother

\confheader{%
	Accepted and scheduled to be presented in the 45th International Conference on Acoustics, Speech, and Signal Processing (ICASSP 2020).
}

\begin{document}
%
\title{The Effect of Data Augmentation on Classification of Atrial Fibrillation in Short Single-Lead ECG Signals Using Deep Neural Networks}



%
\author{\IEEEauthorblockN{Faezeh Nejati Hatamian\IEEEauthorrefmark{1},
Nishant Ravikumar\IEEEauthorrefmark{3},
Sulaiman Vesal\IEEEauthorrefmark{2}, 
Felix P. Kemeth\IEEEauthorrefmark{1},
Matthias Struck\IEEEauthorrefmark{1} and
Andreas Maier\IEEEauthorrefmark{2}}
\IEEEauthorblockA{\IEEEauthorrefmark{1} Department of Image Processing and Medical Technology, Fraunhofer Institute for Integrated Circuits IIS, Erlangen, Germany}
\IEEEauthorblockA{\IEEEauthorrefmark{2} Pattern Recognition Lab, Department of Computer Science,
	Friedrich-Alexander University Erlangen-Nürnberg, Erlangen, Germany}
\IEEEauthorblockA{\IEEEauthorrefmark{3} Centre for Computational Imaging and Simulation Technologies in
	Biomedicine (CISTIB), School of Computing, \\ LICAMM Leeds Institute of Cardiovascular and Metabolic Medicine, School of Medicine, University of Leeds, United Kingdom \\ Email: andreas.maier@fau.de}
}

\maketitle

\begin{abstract}
Cardiovascular diseases are the most common cause of mortality worldwide. Detection of atrial fibrillation (AF) in the asymptomatic stage can help prevent strokes. It also improves clinical decision making through the delivery of suitable treatment such as, anticoagulant therapy, in a timely manner. The clinical significance of such early detection of AF in electrocardiogram (ECG) signals has inspired numerous studies in recent years, of which many aim to solve this task by leveraging machine learning algorithms. ECG datasets containing AF samples, however, usually suffer from severe class imbalance, which if unaccounted for, affects the performance of classification algorithms. Data augmentation is a popular solution to tackle this problem. 

In this study, we investigate the impact of various data augmentation algorithms, e.g., oversampling, Gaussian Mixture Models (GMMs) and Generative Adversarial Networks (GANs), on solving the class imbalance problem. These algorithms are quantitatively and qualitatively evaluated, compared and discussed in detail. The results show that deep learning-based AF signal classification methods benefit more from data augmentation using GANs and GMMs, than oversampling. Furthermore, the GAN results in circa $3\%$ better AF classification accuracy in average while performing comparably to the GMM in terms of f1-score.
\end{abstract}


%
\IEEEpeerreviewmaketitle

\section{Introduction}

Based on a report released by the World Health Organization (WHO), ischemic heart disease, also known as coronary artery disease, were globally acknowledged as one of the most common causes of death in 2016~\cite{WHO}. 
Previous studies have shown that AF increases the risk of embolic stroke by five times~\cite{reiffel2014atrial}. Moreover, detection of atrial fibrillation (AF) in the asymptomatic stage can help prevent strokes and aid in the timely identification of patients that are likely to benefit from anticoagulant therapy~\cite{benjamin2018heart}~\cite{christiansen2016atrial}.

In a healthy heart, the sinoatrial (SA) node, which works as the natural pacemaker, generates electric signals in a periodic pattern. These signals result in the periodic contraction of the heart's atria and ventricles. 
These consecutive contractions translate into the PQRST peaks in an electrocardiogram (ECG) signal. 
Malfunction in the SA node may cause the atria to quiver instead of performing a complete contraction. This can potentially result in blood remaining in the atrial chamber, followed by the formation of clots, and ultimately leading to ischemia and strokes~\cite{levy1998atrial},~\cite{padmavathi2015classification}. 
AF is reflected in the missing of the P-peak and its substitution with an inconsistent and chaotic fibrillatory wave (f-wave) and an irregular R-peak to R-peak (R-R) interval in an ECG signal~\cite{levy1998atrial}.

The existence of a distinct pattern in ECG signals with AF disorder makes machine learning an appropriate tool for the task of distinguishing the normal vs. AF signals. 
The possibility of automatically detecting AF in ECG signals along with the potential clinical impact of doing so, have led to the PhysioNet/CinC challenge organized in 2017, which focused on the classification of ECG signals with AF anomaly~\cite{goldberger2000physiobank, clifford2017af}.
Numerous studies have adopted the traditional machine learning pipeline using handcrafted features and reported models with accurate results. In these approaches, expert knowledge of cardiologists was taken into account to calculate representative and distinctive features. 
These works either focus on detecting the presence of the f-wave instead of the P-peak~\cite{garcia2016application} or are based on anomaly detection in the R-R intervals in ECG signals~\cite{linker2016accurate}.

Hand crafting descriptors and engineering features, however, are very expensive, time-consuming and tedious. Moreover, they demand the availability of expert knowledge. Convolutional Neural Networks (CNNs) provide the means to automate this feature extraction process without the need for expert domain knowledge. Moreover, the hierarchical data processing nature of CNNs produces highly descriptive and informative features. Deep CNNs have shown competitive performance in different computer vision and medical applications and in some cases matched or even surpassed the human accuracy~\cite{goodfellow2016deep}.

Deep CNNs have also been adopted for ECG signal classification. 
Rajpurkar et al.~\cite{rajpurkar2017cardiologist} proposed a 34-layer CNN for classifying 12 cardiac disorders by mapping the ECG signals to their corresponding rhythmic disorder classes. 
Andreotti et al.~\cite{andreotti2017comparing} used ResNets~\cite{he2016deep} for AF detection in ECG signals. The residual connections in such networks help tackle the issue of vanishing gradients, enabling deeper networks to be trained. 
An alternative to the use of CNNs for supervised feature learning, and the use of hand-crafted features, was proposed in~\cite{al2016deep}. In the representation block, they feed the raw ECG signals to a stacked denoising autoencoder and train it in an unsupervised manner. The result is then fed into a softmax regression layer.
Authors in~\cite{isin2017cardiac} use the famous AlexNet for feature extraction. They fed AlexNet with R-T segments, which are extracted from the ECG signals, and direct the result into a 2-layer feed-forward neural network for detecting normal beats, paced beats and Rbbb (Right Bundle Branch Blocks) beats.
%
The work of Ziehlmann et al.~\cite{zihlmann2017convolutional} achieved one of the best classification performances in the PhysioNet 2017 challenge~\cite{clifford2017af}. Their best model is a 24-layer CNN followed by an LSTM. The input to the network is the logarithmic spectrogram of the ECG signals. 
In a similar work, Chauhan et al.~\cite{chauhan2015anomaly} proposed to stack multiple LSTM layers for the task of cardiovascular anomaly detection from ECG signals.

Previous studies on ECG signal classification using deep learning focus on proposing new and powerful architectures that can better handle the aforementioned task. However, to our knowledge, there is neither a universal agreement nor a comprehensive and thorough study on the most effective data augmentation algorithm to be employed to address the issue of the class imbalance. In this work, we aim to close this gap. 
The rest of this paper is organized as follows: 
In section~\ref{sec:framework_TB} we go over our classification pipeline, discuss our contribution and review the theoretical background of the algorithms we employ. Section~\ref{sec:evaluation} is dedicated to describing the experimental setup and summarizing the results. Finally, section~\ref{sec:conclusion} concludes the paper.

\section{Processing Framework and Theoretical Background}\label{sec:framework_TB}
\subsection{Classification Pipeline}
The majority of previous studies employing deep learning for ECG signal classification, follow a common pipeline~\cite{zihlmann2017convolutional, chauhan2015anomaly}: First the 1-D signals are pre-processed. The ECG signals can be segmented into equal-size segments. Moreover, it is common to transform the 1-D signals to 2-D logarithmic spectrograms. In the next step, various data augmentation schemes are typically employed to reduce class imbalance. It is worth noting that not all studies employ data augmentation. Finally, the pre-processed and augmented data is fed into the deep network architecture. 

In this work, we focus on the data augmentation block and evaluate various data augmentation schemes, in order to identify the most suitable ones to enhance classification performance. In the literature and within this community, oversampling is the most common data augmentation algorithm. 
In this paper, we introduce and study the effectiveness of two more augmentation algorithms, namely Gaussian Mixture Model (GMM) and Generative Adversarial Networks (GANs).

\subsection{Gaussian Mixture Model (GMM)}
\subsubsection{Gaussian Probability Distribution Function}
Many random events in nature follow Gaussian probability distribution functions (pdf).
A Gaussian pdf is parameterized by two parameters, i.e., the mean and the variance.
For one dimensional data $\boldsymbol{X}$, the Gaussian pdf is represented by $\mathcal{N}(\mu, \sigma)$ as in Eqn.~\ref{eqn:GaussianPDF1D}.
\begin{equation}\label{eqn:GaussianPDF1D}
\mathcal{N}(\boldsymbol{X}|\mu,\sigma) = \frac{1}{\sqrt{2\pi\sigma^2}} \exp \bigg( -\frac{(\boldsymbol{X}-\mu)^2}{2\sigma^2}\bigg).
\end{equation}

However, for $d$-dimensional data $\boldsymbol{X}=[\boldsymbol{X}_1,...,\boldsymbol{X}_d]$, the mean is a vector $\boldsymbol{\mu}$ and the variance becomes a covariance matrix $\boldsymbol{\Sigma}$. The multivariate Gaussian pdf in this case is mathematically expressed as in Eqn.~\ref{eqn:GaussianPDF2D}.
\begin{equation}\label{eqn:GaussianPDF2D}
\mathcal{N}(\boldsymbol{X}|\boldsymbol{\mu},\boldsymbol{\Sigma}) = \frac{1}{(2\pi)^\frac{d}{2}|\boldsymbol{\Sigma}|^\frac{1}{2} } \exp \bigg( -\frac{1}{2} (\boldsymbol{X}-\boldsymbol{\mu})^T \boldsymbol{\Sigma}^{-1}(\boldsymbol{X}-\boldsymbol{\mu})\bigg) \enspace,
\end{equation}
where $|\boldsymbol{\Sigma}|$ represents the determinant of the covariance matrix. Thus, for this pdf to be well defined, the covariance matrix $\boldsymbol{\Sigma}$ needs to be positive-definite.

\subsubsection{GMM}
Despite the diversity of the Gaussian pdf, not all the processes follow the Gaussian distribution. Such arbitrary data distributions can, however, be approximated using a GMM, i.e., by the weighted sum of more than one Gaussian component, as in Eqn.~\ref{eqn:GMM}:
\begin{equation}\label{eqn:GMM}
p(\boldsymbol{X}|\boldsymbol{\mu}_i, \boldsymbol{\Sigma}_i, w_i) = \sum_{i=1}^{c} w_i \mathcal{N}(\boldsymbol{X}|\boldsymbol{\mu}_i,\boldsymbol{\Sigma}_i) \enspace,
\end{equation}
where $\boldsymbol{\mu}_i$, $\boldsymbol{\Sigma}_i$ and $w_i$ represent the mean, covariance and mixture weight of the $i$-th component. Furthermore, mixture weights $w_i$ need to sum up to one. 
Given the data $\boldsymbol{X}$, the parameters of a GMM are usually estimated via the expectation maximization (EM) algorithm in an iterative manner. 

\subsection{Generative Adversarial Networks (GANs)}\label{sec:GANs}
\begin{figure}[tbp]
	\begin{minipage}[b]{1\linewidth}
		\centering
		\centerline{\includegraphics[width=1\linewidth]{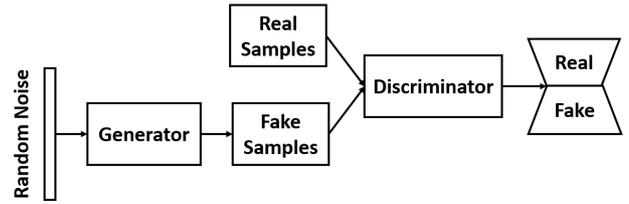}}
	\end{minipage}
	\caption{High level GAN architecture.}
	\label{fig:GAN}
\end{figure}

Generative adversarial networks (GANs)~\cite{goodfellow2014generative} 
are composed of two sub-networks, i.e., a generator and a discriminator. The generator takes a random noise vector $z$ initially as input, which represents the latent space to be learned, and learns the mapping between this latent vector $z$ and the data distribution. The discriminator plays the role of an investigator, who determines the legitimacy of the data generated by the generator from the learned latent representation. The main objective of the generator is to fool the discriminator. These two networks are trained alternatively, which makes them grow together and therefore, compete. The high level GAN architecture is depicted in Fig.~\ref{fig:GAN}.

GAN's training, however, strongly depends on the two competing networks, i.e., the generator and the discriminator, and can be highly unstable. Convolutional neural networks (CNNs) have shown great performance in supervised learning~\cite{maier2019gentle}.
Radford et al.~\cite{radford2015unsupervised} proposed to combine CNNs and GANs and called the final framework Deep Convolutional GAN (DCGAN). In particular, they eliminated all the fully connected layers and substituted them by convolutional layers in the discriminator and transposed convolutional layer in the generator. Moreover, max-pooling and upsampling are replaced by strided convolution and strided transposed convolution, respectively. In addition, in the discriminator all the activation functions are LeakyReLU. In the generator, all the activation functions are ReLU except for the last layer which uses tanh as the activation function. Finally, for further stabilizing the training, Batch Normalization (BN) is used after all layers except the output of the generator and the input of the discriminator. 

\section{Evaluation}\label{sec:evaluation}

\subsection{Dataset}
\subsubsection{Data Statistics}\label{sec:DataStatistics}
The data used in this study is publicly available from the PhysioNet/CinC challenge 2017~\cite{goldberger2000physiobank, clifford2017af}. It comprises $8528$ single-lead ECG signals in four classes, namely $5154$ Normal (N), $771$ atrial fibrillation (AF), $2557$ Noisy and $46$ Other rhythm signals. Following the main objective in this work, we kept the Normal and AF classes and discarded the Noisy and Other rhythm ones. 
The ECG signals' recording times vary between $9$ and $61$ seconds, which are sampled at the rate of $300$ Hz. 
The majority of the signals in the provided dataset is $30$s long, while a smaller subset is $60$s long.

\subsubsection{Data Segmentation}
As different ECG signals in the dataset have different length and the majority of deep classifiers require the samples to have equal size, it is common to segment the signals into smaller segments with equal sizes~\cite{warrick2017cardiac, hong2017encase}. In order to do so, a segment length (SL) is chosen. 
In particular, ECG signals longer than the chosen SL are cropped, whereas signals shorter than the SL are discarded. In this work, we follow~\cite{xiong2017robust} and use SL=$1500$ being equivalent to $5$ seconds.

\subsubsection{Class Imbalance and Data Augmentation}

Using SL=$1500$, AF class has $4674$ signals and Normal class has $31939$ signals that is seven times the AF class, which indicates the presence of significant class imbalance. This is undesirable in supervised learning as it would push the classifier towards being biased on the larger class.
In order to tackle this problem, we use data augmentation on the AF class.
The data augmentation algorithms that we use in this paper are oversampling, GMM and DCGAN and we compare them with no data augmentation, referred to as None. When using GMM and DCGAN for data augmentation, we follow Davari et al.~\cite{davari2019vb}. That means, we train these generative models using only the AF class samples, draw virtual samples from the trained models and combine them with the original samples so that the number of AF class samples equals the Normal class samples.

For training the DCGAN and GMM, we tried both the 1-D and 2-D (logarithmic spectrogram) variants of the ECG signals. DCGAN performed better when trained using the 2-D spectrograms, while GMM performed better when trained on the 1-D signals. The GANs have been shown to have high capacity in producing high quality images~\cite{radford2015unsupervised}. However, GMMs are unable to accurately capture spatial correlations in 2-D images and are used to generate 1-D signals~\cite{davari2019vb}.

\subsubsection{Division of Data into Train, Validation and Test Subsets}\label{sec:TrainTestVal_Division}
In order to divide our data into training set, validation set and the test set, we opt for the following policy: We take $20\%$ of whole dataset as the test set. Then we divide the remaining $80\%$ into $90\%$ training and $10\%$ validation. 
We want to preserve the original data statistics and keep the imbalanced class ratio during the division. Therefore, in order to take $x\%$ from the data, we take $x\%$ from the AF class and $x\%$ from the Normal class and then concatenate them.

\subsubsection{2-D Data Representation}
In order to convert our 1-D signals to 2-D using logarithmic spectrogram, we use the \emph{signal.Spectrogram} module from the Scipy toolbox~\cite{ScipyBib} in python. As for the spectrogram's hyper-parameters, we follow Zihlmann et al.~\cite{zihlmann2017convolutional}, i.e., Tukey window of length $64$ and hop length of $32$ (i.e., $50\%$ window overlap), and shape parameter of $0.25$. 

\subsection{Deep Learning Architectures}\label{sec:architectures}
\subsubsection{Classification Architectures}
\begin{figure}[tbp]
	\begin{subfigure}[b]{.5\linewidth}
		\centering
		\centerline{\includegraphics[width=1\linewidth]{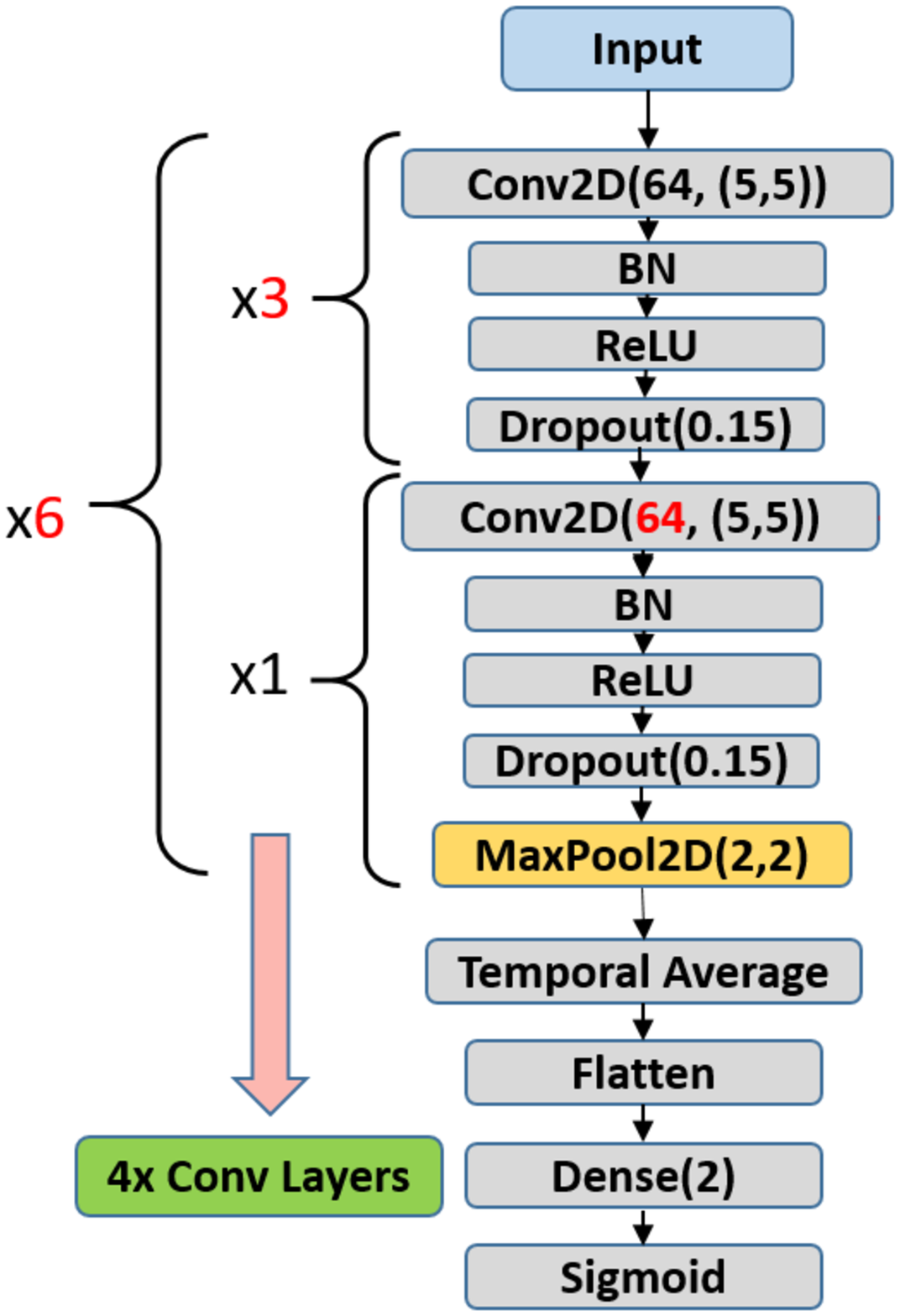}}
		\caption{SOA\_CNN}
		\label{fig:SOA43_CNN_2D}
	\end{subfigure}
	\begin{subfigure}[b]{.5\linewidth}
		\centering
		\centerline{\includegraphics[width=1\linewidth]{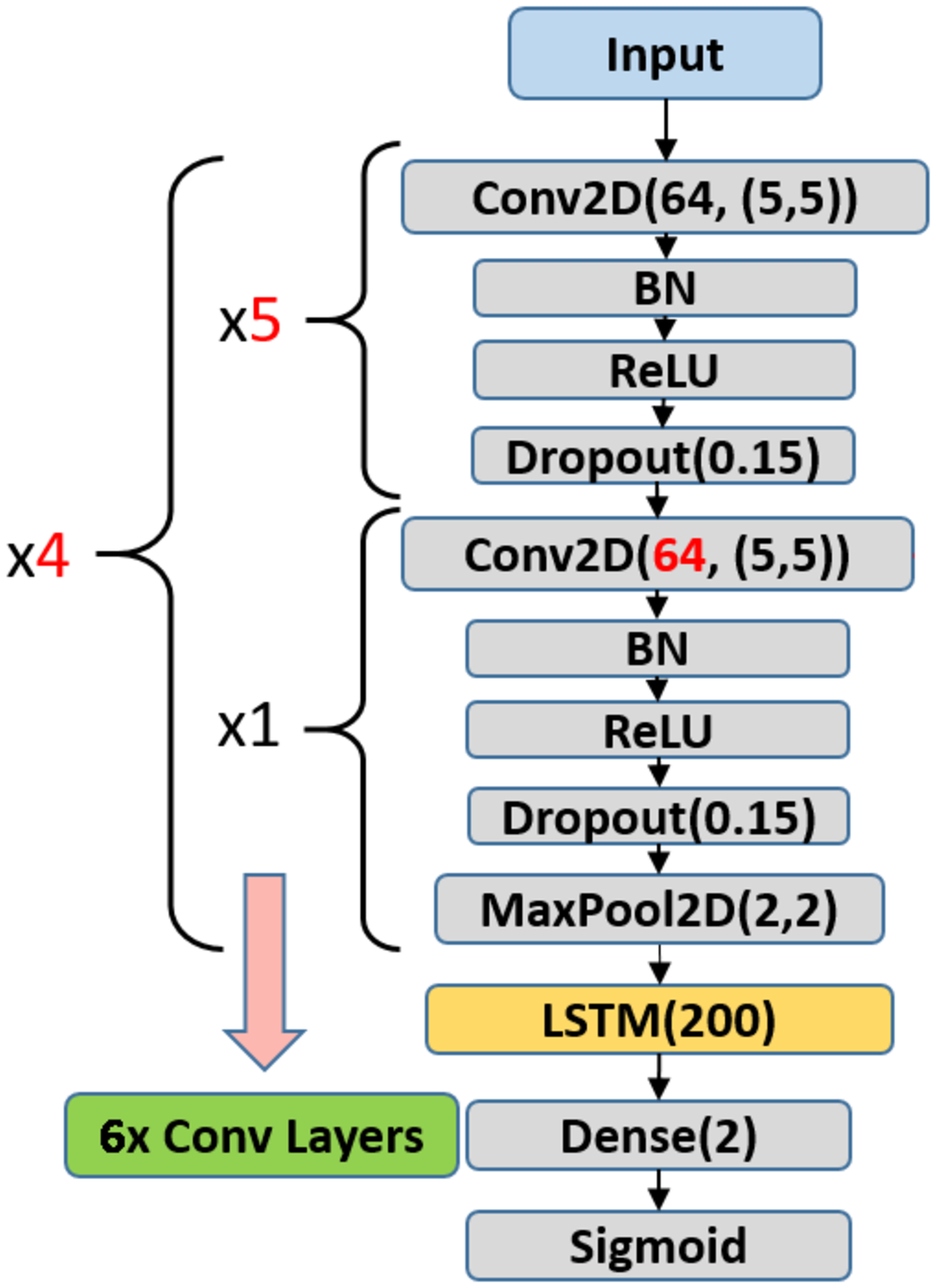}}
		\caption{SOA\_CRNN}
		\label{fig:SOA43_CRNN}
	\end{subfigure}
	\caption{Zihlmann et al.~\cite{zihlmann2017convolutional} architectures: (a) SOA\_CNN, (b) SOA\_CRNN.}
	\label{fig:SOA43_family}
\end{figure}

For the evaluation purposes in this paper, we use one of the leading deep learning-based works in the PhysioNet challenge 2017, by Zihlmann et al.~\cite{zihlmann2017convolutional}. They proposed two different architectures, which are all adapted towards the usage of 2-D spectrogram representation of the ECG signals. SOA\_CNN, which is depicted in Fig.~\ref{fig:SOA43_CNN_2D}, comprises six convolutional blocks. Each block consists of four sub-blocks with a convolutional layer with $5 \times 5$ kernel size, then a batch normalization followed by ReLU activations and a dropout layer. The last sub-block contains a max-pooling layer. The first three convolutional layers have $64$ filters while, the last convolutional layer starts with $64$ filters in the first block, with the filter number being increased by $32$ for each subsequent block.
The output of the convolutional blocks is then averaged on the temporal axis, flattened and fed into a sigmoid classifier. 

Their next proposed architecture, SOA\_CRNN, is very similar to SOA\_CNN. Their differences boil down to the usage of $4$ convolutional blocks instead of $6$, and the substitution of the temporal averaging layer by an LSTM layer with $200$ neurons. This network is illustrated in Fig.~\ref{fig:SOA43_CRNN}. 
They trained SOA\_CRNN in three phases. In the first phase, they replaced the LSTM layer by temporal averaging, and trained the resulting network. Then, they substituted the temporal averaging with an LSTM layer and trained this layer while keeping the rest of the network frozen. At this point, the convolutional blocks and the LSTM layer are trained independently. In the final training phase, the whole network is trained. This training policy effectively initializes parts of the network far better than random weight initialization. 

\begin{table*}[htb]
	\centering
	\caption{Quantitative results associated with AF class augmentation using different data augmentation algorithms. F1 stands for f1-score, AF stands for AF class accuracy and N represents the Normal class accuracy.}
	\begin{tabular}{@{}l@{}|@{ }c@{ }c@{ }c@{ }|@{ }c@{ }c@{ }c@{ }|@{ }c@{ }c@{ }c@{ }|@{ }c@{ }c@{ }c@{}}
		\hline
		& \multicolumn{3}{c@{ }|@{ }}{None} & \multicolumn{3}{c@{ }|@{ }}{oversampling} & \multicolumn{3}{c@{ }|@{ }}{GMM} & \multicolumn{3}{c}{DCGAN} \\
		\hline
		Classifier & F1[\%] & AF[\%] & N[\%] & F1[\%] & AF[\%] & N[\%] & F1[\%] & AF[\%] & N[\%] & F1[\%] & AF[\%] & N[\%] \\
		\hline
		SOA\_CNN & 84.47 & 75.61 & 99.50 & 85.91 & 79.25 & 99.23 & \textbf{87.16} & 85.67 & 98.40 & 86.77 & \textbf{86.31} & 98.15 \\
		SOA\_CRNN & 86.02 & 81.60 & 98.81 & 85.81 & 80.86 & 98.89 & \textbf{87.79} & 82.67 & 99.17 & 86.90 & \textbf{88.34} & 97.81 \\
		\hline
	\end{tabular}%
	\label{tab:aug_GAN}%
\end{table*}%

\subsubsection{DCGAN Architecture}
\begin{figure}[tbp]
	\begin{subfigure}[b]{.5\linewidth}
		\centering
		\centerline{\includegraphics[width=.7\linewidth]{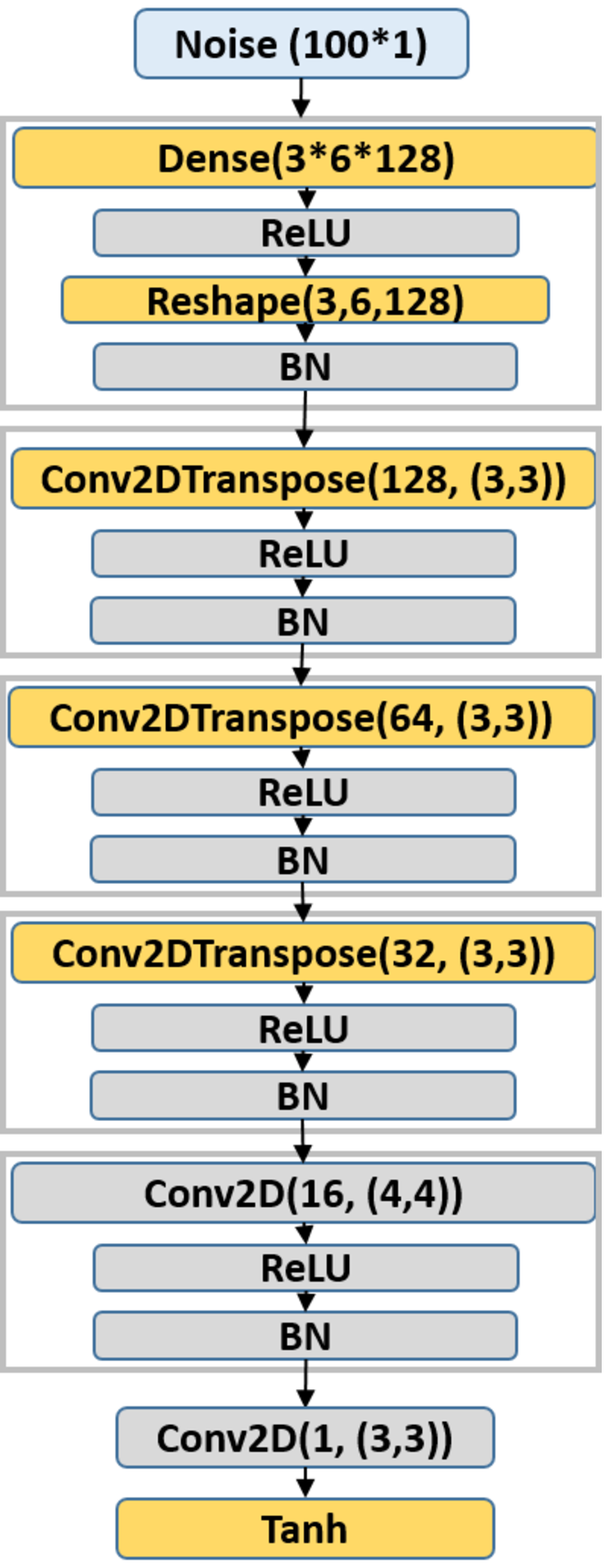}}
		\caption{Generator}
		\label{fig:GAN_gen}
	\end{subfigure}
	\begin{subfigure}[b]{.5\linewidth}
		\centering
		\centerline{\includegraphics[width=.7\linewidth]{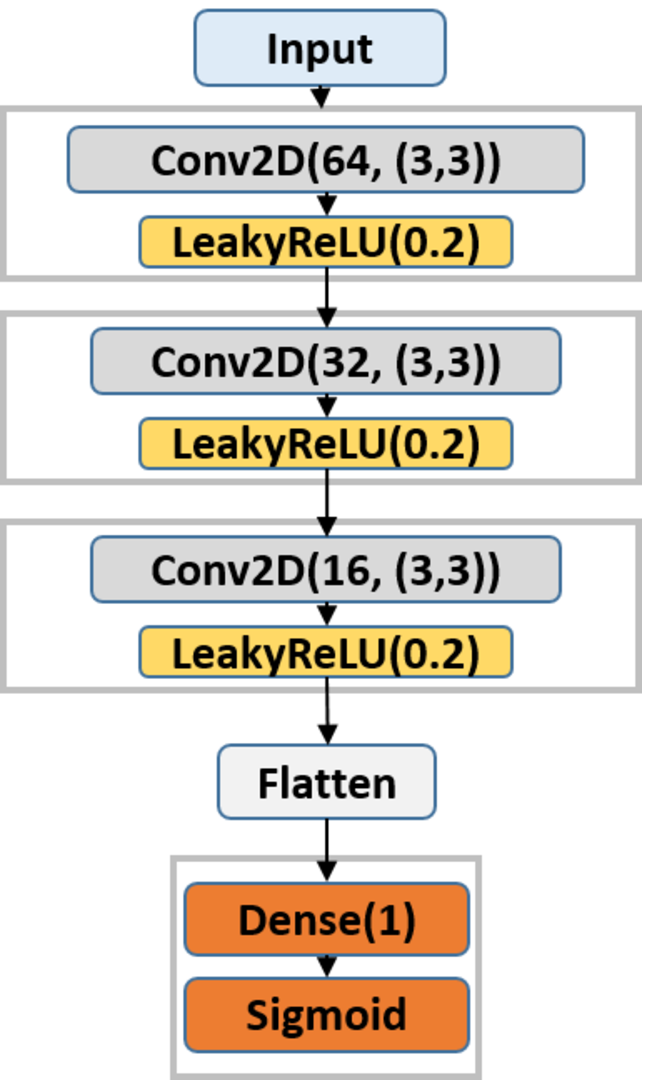}}
		\caption{Discriminator}
		\label{fig:GAN_disc}
	\end{subfigure}
	\caption{The proposed DCGAN generator and discriminator architectures.}
	\label{fig:DCGAN_arch}
\end{figure}

We investigated different architectures for the generator and discriminator in the DCGAN. The best performing combination is depicted in Fig.~\ref{fig:DCGAN_arch}. In our implementation of the DCGAN architecture, we follow the general tips and tricks that are suggested in~\cite{radford2015unsupervised} and explained in Sec.~\ref{sec:GANs}.

\subsubsection{Hyper-parameters}
As preprocessing, we apply mean-std normalization on the data. That means, we subtract the mean of the training data from all data subsets, i.e., training, validation and test subsets. Then we divide the result by the standard deviation (std) of the training data. The mean of the resulting data is zero and standard deviation is one.
For training the deep neural networks, we use the ADAM optimizer with learning rate equal to $10^{-3}$ and decay rate of $10^{-5}$, and run them for $150$ epochs. However, while training the generator and the discriminator of the DCGAN, we use a learning rate equal to $2\times10^{-4}$, as suggested in~\cite{radford2015unsupervised}. We use binary cross entropy (BCE) loss function for training all networks. As for the weight initialization in the network, we use Xavier uniform initialization technique, also known as Glorot uniform. 

We implemented all the neural network architectures using the Keras toolbox with Tensorflow backend~\cite{chollet2015keras} in Python. Apart from the previously explained hyper-parameters, we used the default hyper-parameters of Keras. We refer the reader to the Keras documentation~\cite{chollet2015keras} for detail set of parameter values for reproduction. Furthermore, in the network architecture visualizations in this work, we use the Keras syntax. For example for a 2-D convolution, we may write Conv2D($a$,($b$,$c$)), which $a$ represents the number of convolutional filters and ($b,c$) represents the convolutional kernel size. 

\subsection{Experimental Results}\label{sec:results}
Table~\ref{tab:aug_GAN} presents the classification results that are computed using different data augmentation algorithms, namely oversampling, GMM and DCGAN. For the sake of comparison, we report the results without augmentation (None) as well. It can be observed that using data augmentation always outperforms no augmentation with respect to f1-score. 
Another interesting observation is that in the majority of the cases, applying augmentation on the AF class results in an improvement in both the AF accuracy and the total f1-score. However, in some cases the Normal class accuracy slightly deteriorates ($\approx 1\%$).
Among the augmentation algorithms investigated in this study, the lowest improvement in performance is achieved by oversampling, while GMM and DCGAN enhance the performance the most. Although GMM marginally outperforms DCGAN in terms of the overall f1-score, it turns out that DCGAN results in better AF accuracy while producing comparable Normal class accuracy to GMM.

\begin{figure}[tbp]
	\begin{minipage}[b]{1\linewidth}
		\centering
		\centerline{\includegraphics[width=.9\linewidth]{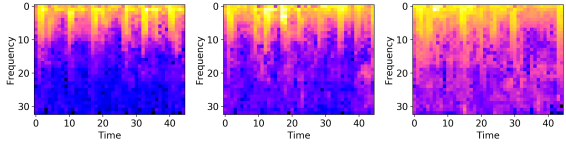}}
	\end{minipage}
	\caption{Sample DCGAN-generated AF ECG spectrograms. The DCGAN is trained on spectrograms of $5$ seconds long signals. Thus, it is expected to see 4-7 peaks.}
	\label{fig:DCGAN_Samples}
\end{figure}

\begin{figure}[tbp]
	\begin{minipage}[b]{1\linewidth}
		\centering
		\centerline{\includegraphics[width=1\linewidth]{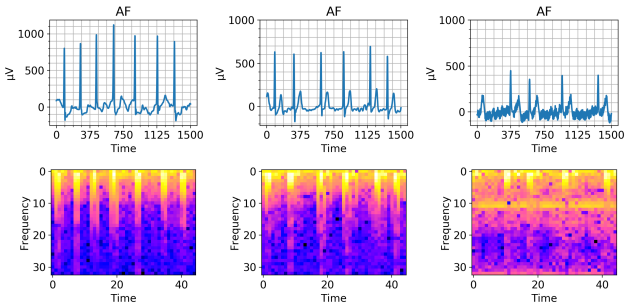}}
	\end{minipage}
	\caption{Original AF signals from the training set (first row) and their corresponding logarithmic spectrogram (second row).}
	\label{fig:orig_AF_spectrograms}
\end{figure}

Figure~\ref{fig:DCGAN_Samples} depicts some spectrograms that are generated by the DCGAN's generator. For the sake of visual comparison, Fig.~\ref{fig:orig_AF_spectrograms} illustrates three real AF ECG signals and their corresponding logarithmic spectrograms.

\begin{figure}[tbp]
	\begin{minipage}[b]{1\linewidth}
		\centering
		\centerline{\includegraphics[width=1\linewidth]{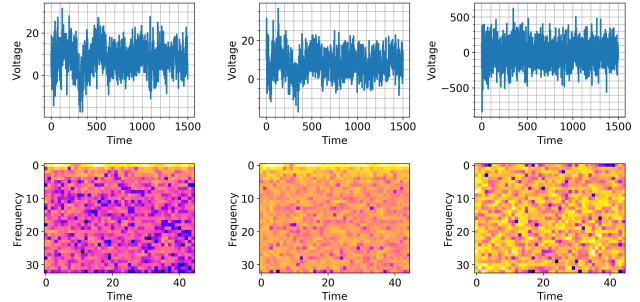}}
	\end{minipage}
	\caption{Sample GMM-generated AF ECG spectrograms.}
	\label{fig:GMM_Samples}
\end{figure}

Figure \ref{fig:GMM_Samples} illustrates the logarithmic spectrogram of two samples that are generated by a GMM, which is trained on the AF class. As it can be observed, these samples hardly resemble an ECG signal visually. However, as GMM resulted in high classification performance improvement, it certainly should have learned the underlying distribution of the AF class distinctive characteristic.   
\section{Conclusion and Discussion}\label{sec:conclusion}
Data augmentation is a popular technique to solve the data imbalance problem in supervised learning.
In this work, we have investigated the impact of various data augmentation algorithms, i.e., oversampling, GMM and DCGAN, on the short single-lead ECG signal classification into AF and Normal classes. The results show that in all cases, data augmentation comes with performance improvement. Oversampling performs comparable to, if not better than, using no augmentation. We should state that we did not expect oversampling to significantly increase the performance because it does not add extra information to the dataset as it simply is a duplication of data. However, it still balances the classes and prevents the classifier to be biased towards the bigger class. On the other hand, the deep learning-based methods seem to really benefit from the DCGAN and GMM. Further, GAN results in better AF accuracy while GMM results in better f1-scores using both classification networks investigated. Both GMM and DCGAN augmentation schemes perform comparably in terms of the Normal class accuracy.

In some cases, using GAN and GMM to augment the AF class causes slight deterioration of the Normal class accuracy. It would be interesting to tackle this problem by applying the learning-based augmentation algorithms, e.g., GMM or GAN, on the Normal class as well. In this manner, the underlying distribution of the Normal class can be learned and this information can be included in the data and exploited by the classifier. Furthermore, other more recent variants of GAN, e.g., improved Wasserstein GAN, outperform DCGAN. In our future work, we intend to conduct a thorough study on the effect of different GAN variants on this application.



\bibliographystyle{IEEEbib}
\bibliography{ref_ICASSP}

\end{document}